\documentclass{PoS}
\usepackage{textcomp}
\usepackage{rotating}

\title{A complementary view of the Galactic plane in TeV gamma rays by HAWC and H.E.S.S. }

\ShortTitle{The Galactic plane in TeV gamma rays by HAWC and H.E.S.S.}

\author{\speaker{Armelle Jardin-Blicq}\\
        Max Planck Institute for nuclear physics, Heidelberg\\
        E-mail: \email{armelle.jardin-blicq@mpi-hd.mpg.de}}

\author{Vincent Marandon\\
        Max Planck Institute for nuclear physics, Heidelberg\\
        E-mail: \email{vincent.marandon@mpi-hd.mpg.de}}
        
\author{Fran\c{c}ois Brun\\
        CEA Paris-Saclay\\
        E-mail: \email{francois.brun@cea.fr}}
\author{for the HAWC\thanks{A full author list can be found here: https://www.hawc-observatory.org/collaboration/icrc2019.php}~  and HESS collaborations\footnote{for collaboration list see PoS(ICRC2019)1177}}

\abstract{
The H.E.S.S. collaboration recently published its Galactic Plane Survey, which is a deep survey of half of the Milky Way in TeV gamma-rays, using almost 10 years of observations~\cite{HGPS}. It contains 78 sources, including 16 new sources. On the other hand, the HAWC collaboration also published its 2HWC catalogue based on 507 days of data~\cite{HAWC_cataolg}. It is the result of the first source search performed with the complete HAWC detector, covering two thirds of the sky. In this search, 39 sources have been reported, amongst them 16 detected for the first time at TeV energies. The HAWC data set doubled since then, revealing 10 more sources. Both instruments are very different in terms of conception, observation and regarding event analysis and reconstruction, but they operate at a similar energy range. In a sense, they are very complementary. I will present a comparison of the Galactic plane seen by both instruments in the overlapping region, highlight the similarities and differences which arise from intrinsic properties of the instruments and from their dedicated data analysis, and show that background estimation is a major ingredient. 
}

\FullConference{36th International Cosmic Ray Conference -ICRC2019-\\
		July 24th - August 1st, 2019\\
		Madison, WI, U.S.A.}

\begin{document}

\section{Instruments and data}
\subsection{HAWC}
The High Altitude Water Cherenkov (HAWC) gamma-ray observatory is a wide field of view instrument located at an altitude of 4100~m in Mexico. Its 300 water tanks, densely packed over an area of 22~000~m$^2$, record the Cherenkov light produced in the water by the secondary charged particles of the atmospheric air shower initiated by gamma rays. 
HAWC is operational since 2015. The study presented here uses 1128 days of data and events falling in the nHit bins 4 to 9, according to the definition given in~\cite{HAWC_crab}, as a compromise between enough statistics and a reasonable PSF - 0.4\textdegree \ for bin 4 down to 0.2\textdegree \ for bin 9. This selection results in a threshold energy of approximately 1 TeV for gamma-rays. The HAWC skymap is shown in the middle of Figure~\ref{allSky}, with a zoom on the Galactic plane. The HAWC sensitivity curve is plotted at the bottom of the same Figure, reflecting the fact that HAWC continuously monitors the sky above it thanks to its 90\textdegree \ field of view, and therefore has an homogeneous observation time smoothly varying as a function of zenith. 

\subsection{H.E.S.S.}
\begin{figure}[ht!]
    \centering
    \includegraphics[width=0.97\linewidth]{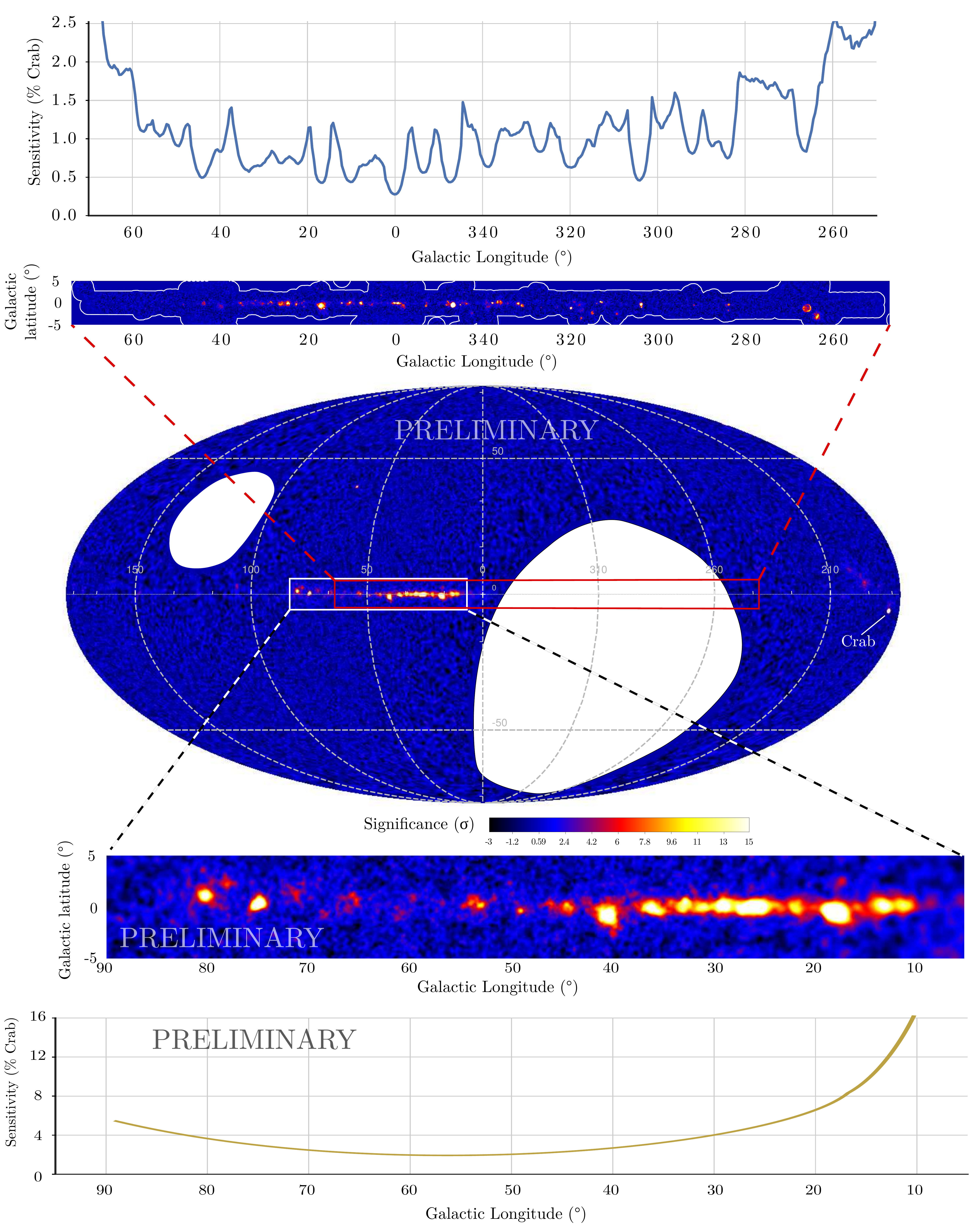}
    \caption[]{\small The map in the middle is a HAWC all-sky significance map for 1128 days of data, using events falling in the nHit bins 4 to 9. A zoom on the Galactic plane is shown below, together with the HAWC sensitivity curve. The red box corresponds to the HGPS significance map with a correlation radius of 0.1\textdegree \ is shown above. At the top, the H.E.S.S. point source sensitivity curve at $b = 0$\textdegree \ is plotted (from~\cite{HGPS}).
    }
    \label{allSky}
\end{figure}

The High Energy Stereoscopic System (H.E.S.S.), located in Namibia at an altitude of 1800~m, is an array of 5~Imaging Atmospheric Cherenkov Telescopes (IACT). Four identical telescopes with a 12~m diameter
tessellated mirror are located on the corners of a square of 120~m length. The fifth one, located at the center has a 32.6~$\times$~24.3~m mirror. They detect simultaneously the Cherenkov light from the air showers produced by gamma rays in the atmosphere. 
The four smallest H.E.S.S. telescopes are taking data since 2003 and surveyed the Galactic plane from 2004 to 2013, producing the H.E.S.S. Galactic Plane Survey (HGPS) with 2700 hours of good quality-selected data~\cite{HGPS}. The fifth one was added in 2012 and did not participate in the survey.
The HGPS significance map is shown at the top of Figure~\ref{allSky} together with the H.E.S.S. point-source sensitivity. It looks much less homogeneous than the HAWC sensitivity curve because H.E.S.S. is a pointing instrument with a defined field of view of 5\textdegree, working only during clear and moonless nights. Hence, it has large differences from a region to another in terms of observation time as well as observation quality due to the evolution of the instrument throughout the survey and to the varying observation conditions from an observation run to another. 
From the HGPS data set, events above 1 TeV are selected to match the HAWC energy range. They are reconstructed using ImPACT~\cite{ImPACT} because of its better performance at the highest energies than the standard reconstruction method using Hillas parameters, used for the HGPS.

\section{Comparison of the Galactic plane by HAWC and H.E.S.S.}
The part of the Galactic plane with longitudes 60\textdegree~<~$\ell$~<~10\textdegree \ where both instruments have reasonable sensitivity  is considered for the comparison.
In this part of the Galactic plane, H.E.S.S. detected 30 sources. For 28 of them, HAWC detected significant emission, even though only 15 sources are actually resolved. Two H.E.S.S. sources are not detected by HAWC: HESS~J1943+213 and HESS~J1911+090. This is actually expected because both of them are below the HAWC sensitivity.
On the other hand, HAWC detected 24 sources and only 15 have a counterpart in the H.E.S.S. catalogue. Hence, 9 sources are detected only by HAWC. As an attempt to understand this discrepancy, the H.E.S.S. map is made as comparable as possible to the HAWC map in terms of angular resolution and map making process.

\newpage

\subsection{Comparable maps in terms of angular resolution}
A main difference between both instruments is their field of view and their angular resolution. While HAWC can observe the whole sky above it, H.E.S.S. is limited to its 5\textdegree \ field of view.
However, with its small field of view comes a good angular resolution. H.E.S.S. has an angular resolution of 0.06\textdegree \ above 1~TeV, almost an order of magnitude better than HAWC. This is clear when looking at the first and second maps of Figure~\ref{GPall}. The HAWC galactic plane map, on the first row, displays more extended emission than the HESS galactic plane map, on the second row. HAWC cannot resolve all the H.E.S.S. sources due to its poorer PSF. However, it is better suited for extended sources. In the spirit of making both maps more comparable, the H.E.S.S. map is convoled with a top hat function of 0.4\textdegree \ radius, the size of the HAWC PSF. The output is the third map of Figure~\ref{GPall}, which looks already much closer to the HAWC map. 

\subsection{Comparable maps in terms of background estimation method}
Another important difference lies in the map making process. In particular, the background estimation methods are adapted to the type of instrument. The ring background method used to make H.E.S.S. maps~\cite{HGPS} defines a ring for each pixel in the field of view where the number of counts outside exclusion regions is used to estimate the background. For the analysis presented here, the exclusion region is defined by the known H.E.S.S. sources and a band of 2\textdegree \ for -1\textdegree~<~$b$~<~1\textdegree. For the HGPS, the typical size of the inner ring radius is 0.7\textdegree, with constant thickness 0.44\textdegree. The inner radius is enlarged to a maximum outer radius of 1.7\textdegree \ if a large portion of the ring area overlaps with exclusion regions. This method is more suited for isolated point-like sources than extended sources because of the constraints on the ring size due to the H.E.S.S. limited field of view.

The background used to make HAWC map is estimated by the direct integration method~\cite{HAWC_crab}. It consists in building an acceptance map by integrating all the events during periods of 2 hours. From the H.E.S.S. side, the closest method to the direct integration technique is to use the acceptance map as background estimation, also referred to as the field-of-view background method~\cite{HESS_background}. The fourth map in Figure~\ref{GPall} uses this method to estimate the background.

\section{Results}
\begin{figure}
    \centering
    \includegraphics[angle=90,width=0.89\linewidth]{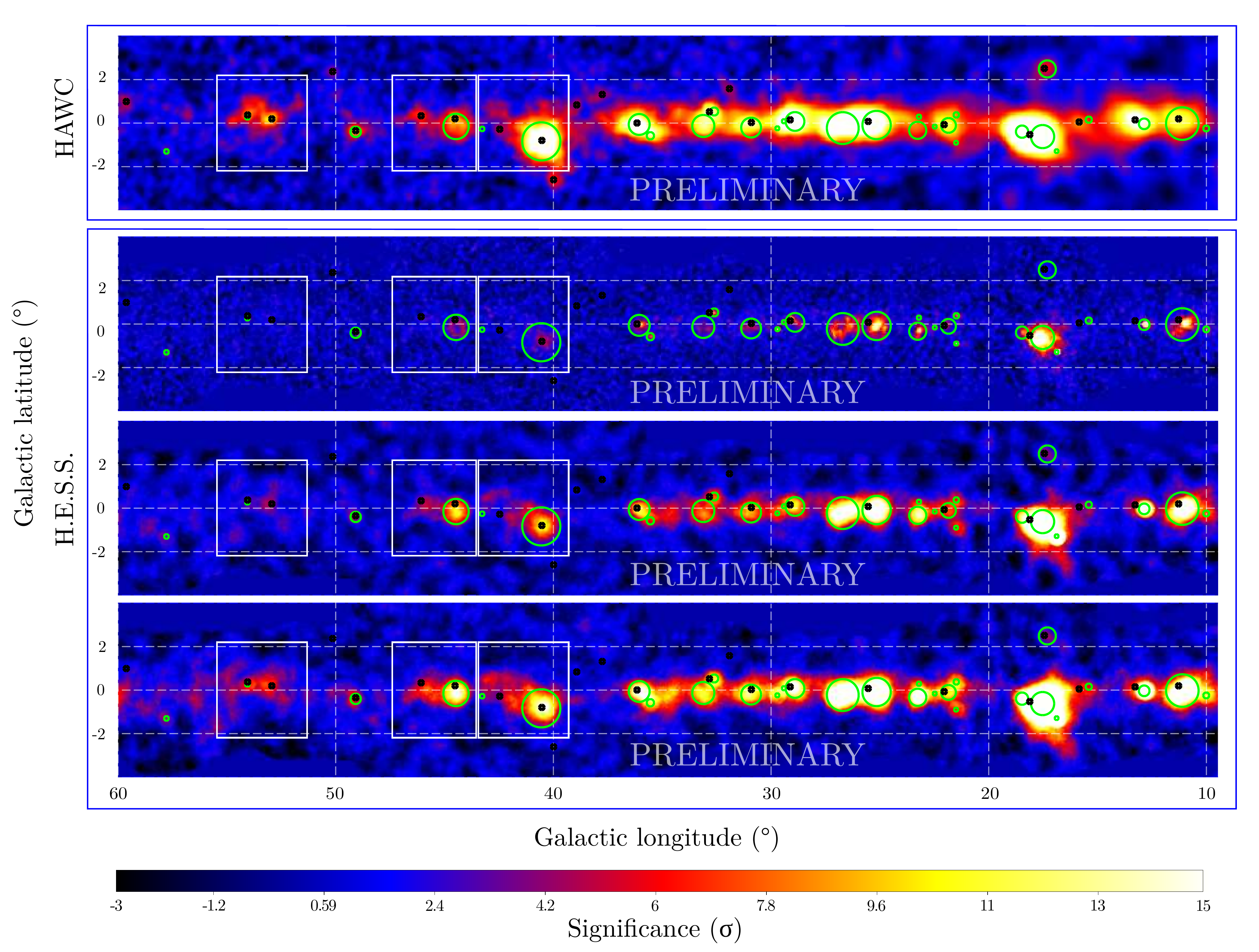}
    \caption[]{\small HAWC and H.E.S.S. galactic plane maps. 
    From top to bottom: \\
     1 - HAWC galactic plane map with 1128 days of data and using events in nHit bins 4 to 9 and the direct integration method for the background estimation. \\
     2 - H.E.S.S. galactic plane map for E > 1 TeV, using ImPACT reconstruction and a correlation radius of 0.1\textdegree; the ring background method is applied on each observation run separately, with an adaptive radius. \\
     3 - Same H.E.S.S. data but using a correlation radius of 0.4\textdegree \\
     4 - Same H.E.S.S. data but using the acceptance map as the background \\
     The green circles are the 68\% containment of the H.E.S.S. sources and the black dots are the location of the HAWC sources. The regions in the white boxes are highlighted in Figure \ref{new_sources}.
    }
\label{GPall}

\end{figure}

\begin{figure}[ht!]
    \centering
    \includegraphics[width=0.8\linewidth]{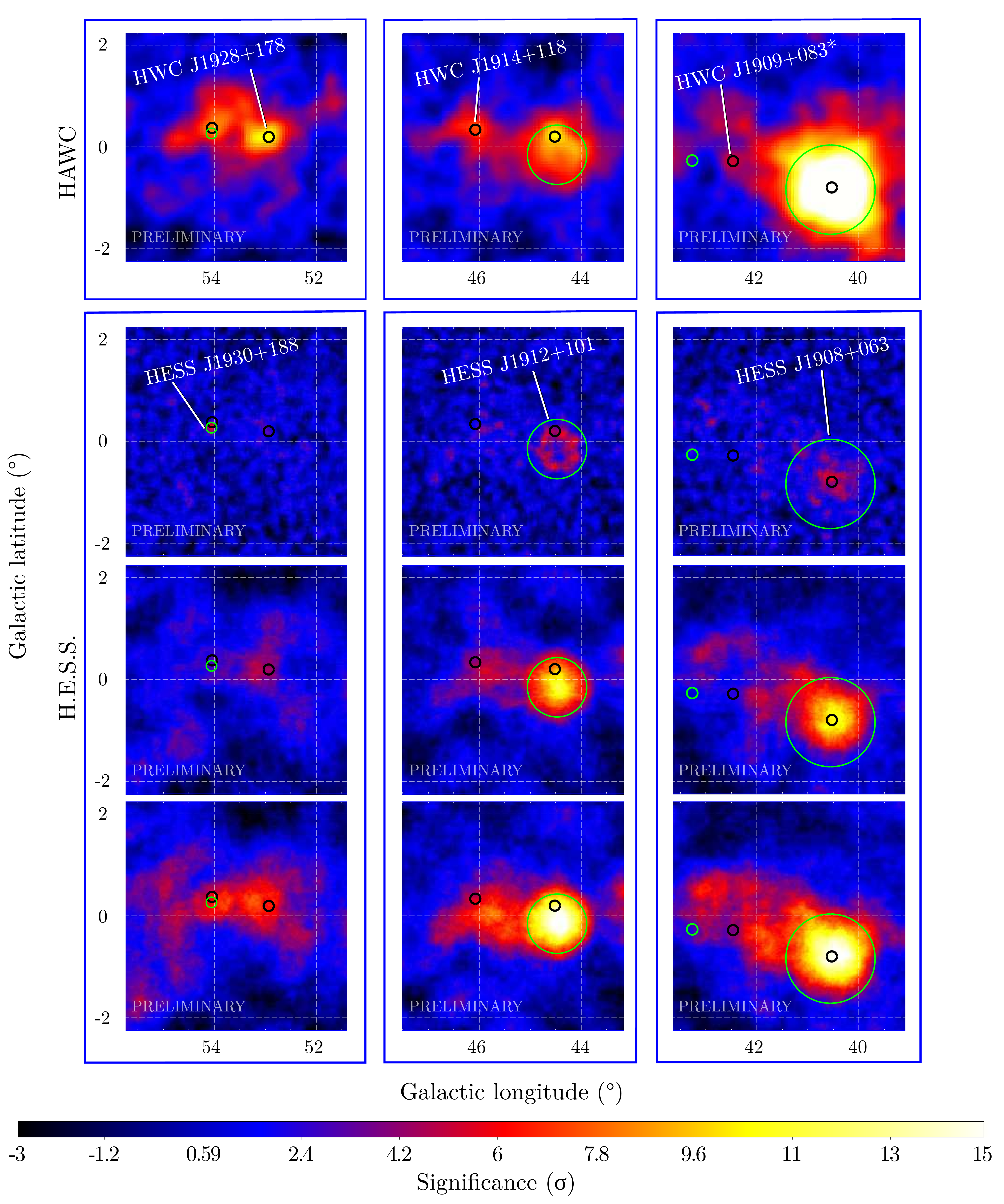}
    \caption[]{\small As for Figure \ref{GPall} but highlighting three specific regions where a HAWC source, undetected by H.E.S.S. (second row) shows significant gamma-ray emission when using a HAWC-like map making process (bottom row).
    }
    \label{new_sources}
\end{figure}

From the 9 sources reported by HAWC and initially not detected by H.E.S.S., 3 of them actually show significant emission above 5$\sigma$ with H.E.S.S. less than half a degree away from a HAWC source. These sources, HAWC~J1928+178, HAWC J1914+118 and HAWC J1909+083* are indicated by the white boxes in Figure~\ref{GPall}, highlighted in Figure~\ref{new_sources}. 
A hint of significant emission near these HAWC sources already appear when using the 0.4\textdegree \ top hat, probably because it integrates more signal. It becomes larger than 5$\sigma$ when using the field of view background method as well.
This shows that, unlike the field of view background method, the H.E.S.S. ring background method tends to remove large scale emission around sources possibly including part of the source itself if extended, wrongly interpreted as background signal. The non detection of the 6 other HAWC sources is consistent in terms of sensitivity as long as sources are more extended than 0.1\textdegree. 

\newpage 

\section{Conclusion}
After understanding the differences between HAWC and H.E.S.S. maps, and making H.E.S.S. maps in a similar way as the HAWC one, this first robust comparison shows that both instruments agree very well with each other and that they observe the same TeV sky. 

A map making procedure developed for a wide field of view instrument is applied to the data taken with an IACT with a small field of view: the angular resolution are made similar by convolving the H.E.S.S. data with a correlation radius of 0.4\textdegree \ and the field of view background method is used for the first time to produce a H.E.S.S. galactic plane map in a comparable way as HAWC. It gives a new point of view on these data: 3 HAWC sources, that were previously undetected by H.E.S.S. show now emission with more than 5$\sigma$ significance in the H.E.S.S. map. This technique seems promising to detect extended sources, eventhough a closer look needs to be taken at the systematics of this approach for very extended sources. 
More observation time could be needed on these 3 regions, and a more detailed morphological and spectral analysis will be done in a future work. 


\end{document}